\documentclass[lettersize,journal]{IEEEtran}
\usepackage{amsmath,amsfonts}
\usepackage{algorithmic}
\usepackage{algorithm}
\usepackage{array}
\usepackage[caption=false,font=normalsize,labelfont=sf,textfont=sf]{subfig}
\usepackage{textcomp}
\usepackage{stfloats}
\usepackage{url}
\usepackage{verbatim}
\usepackage{graphicx}
\usepackage{cite}
\usepackage{xcolor}

\usepackage{amsmath}
\usepackage{multirow}
\usepackage{makecell}
\usepackage{threeparttable}
\usepackage{soul}
\setstcolor{black}

\begin{document}

\title{Video Soundtrack Generation by Aligning Emotions \\ and Temporal Boundaries}
\author{
    Serkan Sulun, Paula Viana and Matthew E.~P.~Davies
    \thanks{Serkan Sulun and Paula Viana are with the Institute for Systems and Computer Engineering, Technology and Science (INESC TEC), 4200–465 Porto, Portugal (e-mail: serkan.sulun@inesctec.pt, paula.viana@inesctec.pt). Paula Viana is also with ISEP, Polytechnic of Porto, School of Engineering, 4200-072 Porto, Portugal (e-mail: pmv@isep.ipp.pt). Matthew E. P. Davies is an independent researcher.}
    }

\maketitle

\begin{abstract}
Providing soundtracks for videos remains a costly and time-consuming challenge for multimedia content creators. We introduce \mbox{EMSYNC}, an automatic video-based symbolic music generator that creates music aligned with a video's emotional content and temporal boundaries. It follows a two-stage framework, where a pretrained video emotion classifier extracts emotional features, and a conditional music generator produces MIDI sequences guided by both emotional and temporal cues. We introduce boundary offsets, a novel temporal conditioning mechanism that enables the model to anticipate upcoming video scene cuts and align generated musical chords with them. 
We also propose a mapping scheme that bridges the discrete categorical outputs of the video emotion classifier with the continuous valence-arousal inputs required by the emotion-conditioned MIDI generator, enabling seamless integration of emotion information across different representations. Our method outperforms state-of-the-art models in objective and subjective evaluations across different video datasets, demonstrating its effectiveness in generating music aligned to video both emotionally and temporally. Our demo and output samples are available at \url{https://serkansulun.com/emsync}.
\end{abstract}


\section{Introduction}

The online distribution of user-generated multimedia content is expanding at an exponential rate thanks to affordable, high-quality recording equipment and video editing software~\cite{multimedia}. A key challenge in content creation is providing suitable soundtracks to enhance viewer engagement~\cite{soundtrack}. However, unauthorized use of commercially published music infringes copyright, preventing monetization for creators on platforms like YouTube. Alternatives such as purchasing music, hiring composers, or searching for royalty-free tracks are often costly, time-consuming, or fail to ensure proper synchronization with video content. Automatic video-based music generation offers a promising solution to this problem~\cite{zhuo,di,kang,xmusic}. 

Existing approaches generate music as audio waveforms, which lack editability~\cite{musicgen1,musicgen2} {\color{black}because they collapse composition, performance, and mastering into a single, inseparable signal.} By contrast, generating music in a symbolic MIDI format offers greater flexibility {\color{black}by maintaining these stages as discrete, modifiable events.} MIDI functions as a digital score, encoding instrument names, note pitches, durations, and velocities. {\color{black}Because these elements are editable, musicians can modify} compositions or refine performances in digital audio workstations (DAWs). 

This work focuses on generating music in MIDI format from arbitrary videos using deep neural networks (DNNs). We use ``MIDI" and ``music" interchangeably, as MIDI is our exclusive format. A major challenge in training video-to-MIDI DNNs is the absence of large-scale paired video-MIDI datasets. Existing datasets are either domain-specific, such as pianist hand videos~\cite{sighttosound} or MIDI files with lyrics metadata~\cite{epia}, or contain limited samples, around 1k~\cite{di,zhuo}. To address this, we develop a model that generates music for any video type by leveraging the Lakh MIDI Dataset~\cite{lmd}, the largest available collection with 176,581 samples, to ensure diverse and high-quality outputs.

Since the Lakh MIDI dataset lacks corresponding videos, we adopt a two-stage approach: extracting video features relevant to music generation and using them as conditioning inputs. Drawing from the musicology literature, we identify two essential aspects for effective video-based music generation: emotional alignment~\cite{emotion_motivation1} and temporal synchronization~\cite{temporal_motivation2}. We name our model \textit{EMSYNC}, as it aligns music with video by matching their \textbf{em}otions, and \textbf{sync}hronizing their temporal boundaries. 
While our method is applicable to any selection of temporal boundary and emotion representation, we define musical boundaries using long-duration chords, video boundaries using scene cuts, and represent emotions with a valence-arousal model~\cite{valence_arousal}. Specifically, we extract scene cut locations from the input video and guide the music generator to produce long-duration chords near locations that are rhythmically and harmonically compatible with the rest of the generated music. We also use our pretrained video emotion classifier to estimate discrete emotion probabilities~\cite{vemoclap, trailer}, map them to valence-arousal values, and condition our music generator accordingly.

Temporal conditioning in MIDI-generating models presents unique challenges. While deep transformer models can handle time-based data, such as videos, their sequence dimension correlates linearly with time due to a fixed frame rate~\cite{vivit,swin}. In contrast, MIDI is processed using an event-based representation {\color{black}in modern applications~\cite{di,kang,xmusic}}, where sequence and time dimensions are not linearly correlated~\cite{event_encoding}. In event-based MIDI encoding, two primary token types exist: ``note" and ``time shift". Note tokens represent pitch, while time shift tokens advance the time axis, capturing both note durations and silences. Each time shift token specifies a time increment, forming a one-dimensional sequence where the position in the sequence does not directly correspond to the position in time. The key advantage of event-based encoding is the absence of a fixed time grid, allowing for expressive timing variations that reflect human musicianship. Unlike state-of-the-art video-based music generators that rely on fixed, coarse time grids~\cite{zhuo,di,kang,xmusic}, we introduce a novel temporal conditioning method that preserves event-based encoding, enabling fine-grained temporal control.

The selection of temporal features for video-music synchronization is also non-trivial. Existing work matches video motion speed and motion saliency with musical note density, producing a dense representation of temporal dynamics that leads to continuous conditioning and frequent changes in note density~\cite{di,xmusic,zhuo,kang}. Our inspection of the output samples from this approach identified two key limitations. First, the continuously varying note density can disrupt rhythmic consistency, making the underlying beat unstable and harder to follow. Second, because the mapping is dense, synchronization between music and video is difficult to perceive unless there are strong contrasts in motion speed and note density across consecutive sections. Motivated by works in musicology and soundtrack composition, we choose a temporal representation of video using \textit{sections}, separated by the sparse temporal boundaries of scene cuts~\cite{temporal_motivation1,temporal_motivation2}. 
{\color{black}By focusing on these sparse boundaries rather than continuous video features, the model only needs to align with discrete temporal points instead of constantly modifying note density, therefore avoiding potential rhythmic instabilities. To achieve this, we introduce \textit{boundary offsets}, which are auxiliary inputs for each token representing the time remaining until the next scene cut. This allows the autoregressive generator to anticipate upcoming boundaries at every step and align generated musical events with them, ensuring precise synchronization.} We conduct objective and subjective evaluations to measure the temporal synchronization between input video and generated music, and demonstrate that our approach outperforms dense temporal conditioning.

In addition, we tackle the challenge of using emotions as an intermediary to link music generation with video. Our pretrained video emotion classifier outputs probabilities of categorical emotions \cite{vemoclap}, trained on Ekman‑6, one of the largest video-emotion datasets providing categorical labels~\cite{ekman6}. 
{\color{black}For music, we previously labeled the Lakh MIDI dataset~\cite{lmd} with valence-arousal values~\cite{access}, producing a training set with a sample size 90 times that of the largest emotion-labeled MIDI dataset~\cite{midiemotion1} and a total duration 187 times that of the longest~\cite{midiemotion3}.}
To bridge these different emotion representations, we introduce a mapping between categorical and dimensional emotions using results from prior user studies~\cite{mapping}. This enables the use of larger datasets for improved performance, such as the Ekman-6 video dataset with categorical labels~\cite{ekman6} and our previous MIDI dataset with dimensional labels~\cite{access}, while supporting training on either type of emotion annotation and maintaining flexibility.

We compare our method to state-of-the-art video-based MIDI generators through objective and subjective evaluation on different datasets.
Our main contributions are as follows:

\begin{itemize}

    \item We present a new framework to train a video-based MIDI generator that surpasses state-of-the-art models across all subjective and a majority of the objective metrics for all datasets.
    \item We introduce the concept of boundary offsets for temporal conditioning in transformers, ensuring precise alignment between the input video and the output MIDI.
    \item Based on psychological research{\color{black}~\cite{mapping}}, we propose a mapping between discrete emotion categories and continuous valence-arousal values, enabling the integration of multimodal data labeled with different emotion representations for video-to-music generation.

\end{itemize}

\section{Related work}

While our method applies to arbitrary videos, several works focus on generating symbolic music for specific types of videos, such as those featuring human movements like dancing or instrumental performances. The Foley Music model~\cite{foley} generates MIDI from videos of musicians by processing body keypoint movements using a Graph Convolutional Network~\cite{gcn} and a Transformer~\cite{transformer}. Similarly, Koepke et al.~\cite{sighttosound} and Su et al.~\cite{audeo} use deep neural networks with residual connections to generate symbolic music from videos of finger movements on piano keyboards. Due to their specialized nature, these approaches rely on datasets containing video-MIDI pairs, however, these datasets typically contain fewer than 1k samples~\cite{music_dataset,video_midi_dataset,sighttosound,audeo}. The RhythmicNet model employs a multi-stage process to generate music from dance videos by predicting beats and style, generating a drum track, and subsequently creating multitrack music~\cite{rhythmicnet}.

Some studies explore the more general task of generating symbolic music for arbitrary videos. The Controllable Music Transformer (CMT) generates music based on video features such as motion speed, motion saliency, and timing~\cite{di}. It processes music using an extended Compound Word representation, where each token encodes type, beat/bar marking, note strength, note density, instrument, pitch, and duration~\cite{compound_words}. 

In a follow-up to CMT, Zhuo et al. introduced the V-MusProd model alongside the Symbolic Music Videos dataset, containing video-MIDI pairs~\cite{zhuo}. They sourced MIDI data from piano tutorial videos, automatically transcribing audio using the Onsets and Frames model, resulting in 1140 samples~\cite{onset_and_frames}. Video features such as color histograms, RGB frame differences for motion, and high-level CLIP features~\cite{clip} were extracted to train three modules sequentially: one generating chord sequences, another generating the melody, and a third generating the accompaniment.

The Video2Music model shares similarities with our approach by utilizing both low-level video features and high-level emotional conditioning, but differs in application~\cite{kang}. The authors compiled the MuVi-Sync dataset, consisting of 748 music videos labeled with musical attributes like note density, loudness, chords, and key. Their encoder-decoder transformer takes low- and high-level video features to generate chord sequences. These sequences are then arpeggiated using fixed patterns to create the final MIDI output. However, relying on fixed arpeggiation patterns limits the musical diversity of the generated pieces.

The XMusic model generates MIDI for videos by analyzing the video in terms of emotional and temporal features~\cite{xmusic}. It uses Compound Word representation and performs dense, bar-by-bar conditioning. A ``selector" module trained with human feedback chooses the best music from a batch of output samples in the end. The model is trained on a privately-collected closed-source dataset.

CMT, V-MusProd, Video2Music, and XMusic all use a quantized temporal representation (32nd notes or coarser) and rely on dense temporal conditioning, often treating each musical bar independently. In contrast, our model adopts an event-based representation with independent time-shift tokens at an 8\,ms resolution. This fine-grained resolution captures subtle timing variations, reflecting how humans naturally play music, rather than imposing rigid, quantized beat subdivisions~\cite{event_encoding}. Furthermore, our use of sparse temporal conditioning in absolute time eliminates reliance on musical bars, enabling the model to handle any type of MIDI, including expressive performances that may not follow strict bar structures.

\section{Methodology}

In this section, we present our video-based symbolic music generator. Since there are no datasets where videos are paired with symbolic music, we employ a two-stage approach, i.e., using independent modules for video analysis and conditional music generation, and then combine them. 

In order to make our contributions self-contained, we briefly summarize the existing components, namely, our video emotion classifier and emotion-based MIDI generator from prior work. Next, we describe how we integrate these components through a mapping scheme that bridges different emotion representation modalities, along with our novel temporal conditioning mechanism for generating MIDI that is temporally aligned with the input video.

\subsection{Overview of prior components}

\subsubsection{Video emotion classification}

Our video emotion classifier exploits publicly available pretrained models for a multimodal video analysis and uses the resulting pretrained features to classify the emotions of arbitrary videos~\cite{vemoclap,trailer}. It employs pretrained models for automatic speech recognition (ASR), optical character recognition (OCR), facial expression classification, audio classification, and image understanding.
Textual features obtained from ASR and OCR are further processed using a text sentiment classifier. 
Cross-attention layers~\cite{transformer} integrate and process these multimodal features, and a linear layer produces the final emotion probabilities. 
The model is trained on the Ekman-6 dataset~\cite{ekman6}, consisting of 1637 videos labeled with the six basic emotions derived from the original work of Ekman: anger, disgust, fear, joy, sadness, and surprise~\cite{ekman}. 

\subsubsection{Emotion-based MIDI generation}
\label{sec:emotion_based}

We have previously collected valence and arousal values for songs in the Lakh Pianoroll Dataset \cite{lpd} and built a conditional transformer to use the resulting labeled dataset~\cite{access}. To integrate continuous-valued valence and arousal features with discrete musical note tokens, we projected valence and arousal into a vector space using separate linear layers. Musical input tokens are projected into vectors of the same dimensionality using an embedding layer. These vectors are then concatenated with the projected valence and arousal vectors along the sequence dimension and fed into the transformer body with relative global attention~\cite{musictransformer}. 

\subsection{Proposed method}

\begin{figure}[t] 
    \centering
    \includegraphics[width=0.7\columnwidth]{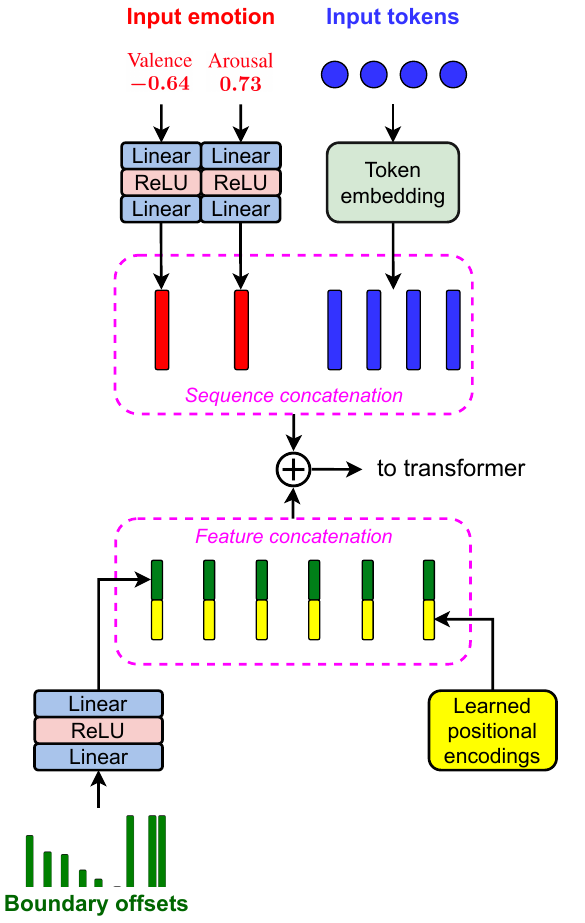}
    \caption{Our music generator. Numbers underneath valence and arousal are sample input values for illustration. {\color{black} Input tokens consist of discrete melodic events (e.g., \texttt{ON}, \texttt{OFF}, and \texttt{TIMESHIFT}) and special functional tokens (e.g., \texttt{START}, \texttt{BAR}, and \texttt{CHORD}). Boundary offsets are a sequence of scalar values, illustrated as the heights of the green bars. The plus sign indicates tensor addition. The dashed magenta boxes indicate concatenation. In sequence concatenation, the encoded valence and arousal vectors with shape (1, D) and the embedded input token sequence with shape (L, D) are concatenated along the first dimension, resulting in a tensor of shape (L+2, D). In feature concatenation, the encoded boundary offsets and positional embeddings, both with shape (L+2, D/2), are concatenated along the second dimension, producing a tensor of shape (L+2, D).}}
    \label{fig:music}
\end{figure}

\begin{figure}[t]
    \centering    \includegraphics[width=0.65\columnwidth]{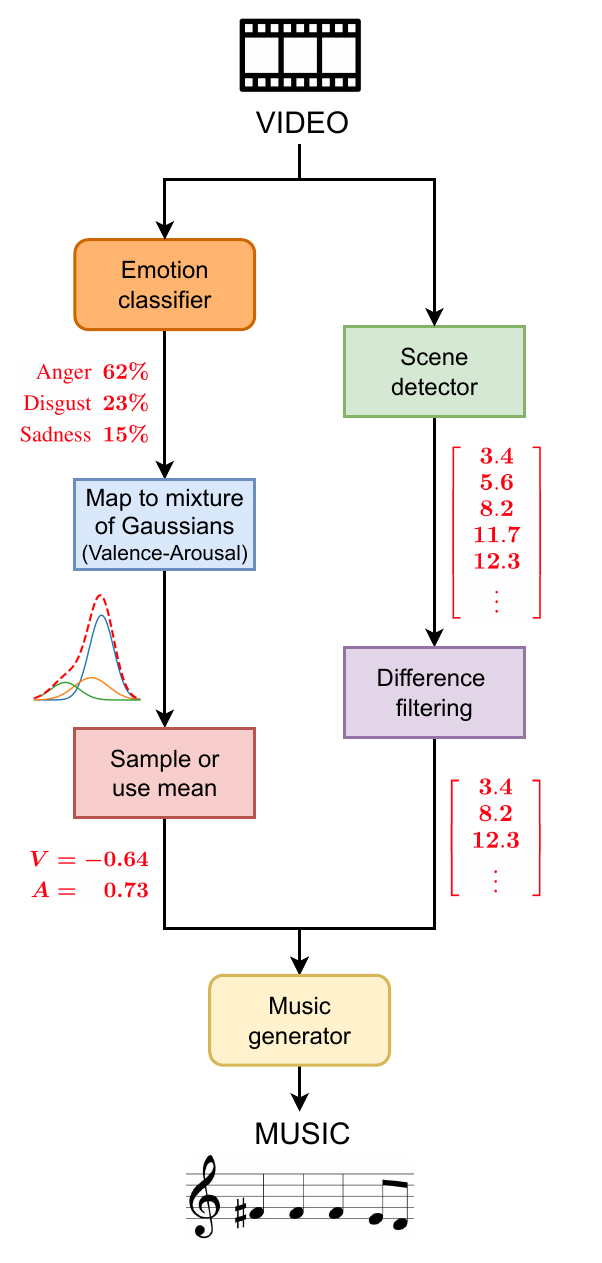}
    \caption{Video-based music generation pipeline. The text and the image next to the arrows demonstrate sample values for illustration.}
    \label{fig:model}
\end{figure}

We extend our previous emotion-based MIDI generator~\cite{access} by introducing temporal boundary conditioning. Figure~\ref{fig:music} illustrates the complete music generation architecture, where the upper section depicts the emotion-conditioning mechanism and the lower section shows the boundary-conditioning mechanism. Through the proposed emotion mapping, we integrate our video emotion classifier with the MIDI generator, forming a unified video-based MIDI generation framework, as shown in Figure~\ref{fig:model}.

\subsubsection{Training dataset and preprocessing}

We train our music generator on the Lakh Pianoroll Dataset (LPD)~\cite{lpd}, which contains 174,154 pianorolls derived from the Lakh MIDI Dataset~\cite{lmd}. We tokenize the pianorolls using an event-based symbolic music representation~\cite{event_encoding}. Specifically, an \texttt{ON} (note on) token marks the start of a note, and an \texttt{OFF} (note off) token marks its end. These tokens also encode pitch and instrument information. For example, a piano note with a MIDI pitch of 60 (C4) is denoted as \texttt{PIANO\_ON\_60}. Since a larger number of instruments increases vocabulary size, we use the Lakh Pianoroll Dataset-5 variant, where all instrument tracks are merged into five predefined categories: bass, drums, guitar, piano, and strings \cite{lpd}. However, our method is adaptable to datasets with different instrument groupings.

\texttt{TIMESHIFT} tokens are used to move along the time axis, representing both note durations and the silences between them. Each token specifies a time increment in milliseconds. For example, an 800\,ms shift is encoded as \texttt{TIMESHIFT\_800}. We use a temporal resolution of 8\,ms with a maximum shift of 1000\,ms. Longer durations are represented using multiple consecutive \texttt{TIMESHIFT} tokens. 
We also use the \texttt{START} tokens to mark the beginning of the songs, the \texttt{BAR} tokens to indicate the musical bars, and the \texttt{PAD} tokens to standardize input sequence lengths in minibatches. We introduce the \texttt{CHORD} token to mark long-duration chords as temporal boundaries, as they often serve as anchor points in musical compositions~\cite{chords}. However, this is a design choice, and our method can accommodate any selection of temporal boundaries such as those separating melodic phrases or song sections.

To fully exploit the available training data, we implement several strategies. Since we aim to generate multi-instrument compositions, we prioritize pieces with three or more instruments. However, one third of the Lakh MIDI dataset contains songs with only one or two instruments, with classical piano pieces dominating the single-instrument subset~\cite{lmd}. As these samples still contain valuable information on musical structure, we avoid filtering them out. Instead, we prepend special tokens: \texttt{FEWER\_INSTRUMENTS} for songs with two or fewer instruments and \texttt{MORE\_INSTRUMENTS} for those with three or more. These tokens allow users to specify instrumentation preferences at inference time while leveraging the entire dataset during training. Additionally, to utilize MIDI samples without valence-arousal labels, we allow the model to accept NaN (Not A Number) values for valence or arousal, enabling both conditional and unconditional generation. If valence or arousal is unspecified, we assign it a NaN value and use a learned vector in place of its projection vector. 

\subsubsection{Temporal boundary matching}
\label{sec:boundary}
\begin{figure}[t] 
    \centering
    \includegraphics[width=0.99\columnwidth]{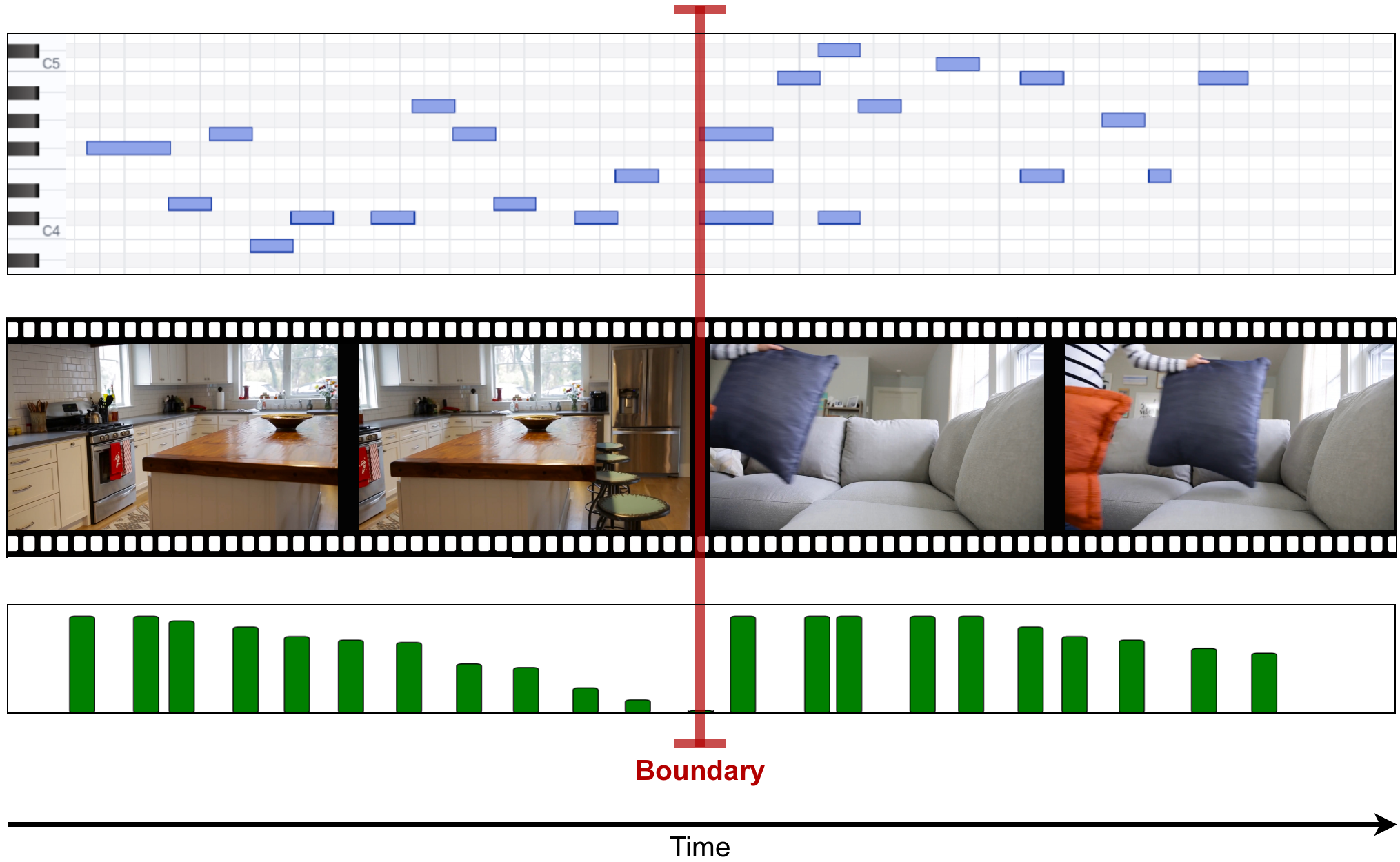}

    \caption{Illustration of boundaries. Top: In symbolic music, a chord with three or more simultaneous notes and a duration exceeding a set threshold defines a musical boundary and is used during training. {\color{black}As \texttt{CHORD} is a special token without duration or pitch, it is omitted from the pianoroll representation. Its temporal location, however, aligns with the red line indicating the boundary.} Middle: In video, scene cuts serve as video boundaries and are used during inference. Bottom: Boundary offset represents the temporal distance to the next boundary. {\color{black}{Each offset is a scalar value, visualized as the height of the corresponding green bar.}} These figures are illustrative; offsets do not perfectly align with the music except at the boundary.}
    \label{fig:boundary}

\end{figure}

Our model generates music that is synchronized with the video's temporal boundaries. We define temporal boundaries as scene cut locations in video and long-duration chord locations in music. Since we employ a hybrid approach, we first train a music generator capable of incorporating boundary locations as input and producing music with chords near these boundaries. To achieve this, we label the chords in the Lakh Pianoroll Dataset. During training, we consider only guitar and piano chords with at least three simultaneous notes that last for a minimum of two beats. These chords are labeled by inserting a \texttt{CHORD} token before the first \texttt{ON} token of each chord. An example of a labeled chord is shown in the top row of Figure \ref{fig:boundary}. 
In training, the temporal locations of chords from the ground-truth MIDI serve as input boundaries, whereas during video-based inference, we replace these boundaries with video scene cut locations.

Chords are integral to a melody, providing harmonic and rhythmic support to surrounding notes, both preceding and following~\cite{chords}. During inference, forcing a \texttt{CHORD} token into the sequence at a specific location may cause the chord to sound off-beat or overly abrupt. This occurs because the model, having no prior knowledge of the upcoming chord, may generate preceding notes that do not align naturally with it. To address this, we develop a method that enables the model to ``anticipate" upcoming chords and generate preceding notes and time shifts accordingly. Additionally, we train the model to generate the \texttt{CHORD} token itself, ensuring rhythmic consistency in the generated music. To achieve this, we define \textit{boundary offsets} for each input token, representing the remaining time until the next boundary. These offsets are capped at a maximum value and, since our music generator is autoregressive, they are computed based on future chords rather than past ones. Sample offsets are illustrated in the bottom row of Figure \ref{fig:boundary}.

We process boundary offsets through a feed-forward network, generating boundary offset encodings, which are then concatenated with learned positional encodings \cite{learned_position} along the feature dimension, as shown in the lower part of Figure \ref{fig:music}. This architectural choice is motivated by several key factors. First, similar to the emotion-conditioning mechanism, we inject boundary offset encodings at the input level rather than within the transformer body, ensuring that the core model remains unchanged. This allows the model to process inputs with or without boundary offsets, enabling seamless fine-tuning by repeating the learned positional encodings along the feature dimension when boundary offsets are absent. Second, we avoid adding boundary offset encodings directly to learned positional encodings to maintain a distinction between the two. Finally, recent studies suggest that decoder-only transformers can implicitly learn positional encodings through their internal weights, even without explicitly adding them~\cite{position1,position2}. Based on this insight, we halve the feature length of learned positional encodings and allocate the remaining feature space to boundary offset encodings. The resulting vector sequence consists of positional encodings augmented with boundary offset encodings. Following the standard transformer model, we add this sequence to the token embeddings~\cite{transformer} before passing it to the transformer body with relative global attention~\cite{musictransformer}. 
The transformer's input is represented as:
{\color{black}

\begin{equation*}
\begin{split}
\mathbf{X}_{\text{trans}} &= 
[\text{FFN}_v(x_v), \, \text{FFN}_a(x_a), \, \text{Embedding}(\mathbf{x}_t)]_1 \\
& \quad + [\text{FFN}_b(\mathbf{b}), \, \mathbf{W}_{\text{pe}}]_2
\end{split}
\end{equation*}
\vspace{-3mm}
\begin{flalign*}
&\text{FFN}_v(x_v), \text{FFN}_a(x_a) \in \mathbb{R}^{1 \times D}, \quad \text{Embedding}(\mathbf{x}_t) \in \mathbb{R}^{L \times D}, \\
&\text{FFN}_b(\mathbf{b}), \mathbf{W}_{\text{pe}} \in \mathbb{R}^{(L+2) \times \frac{D}{2}}, \quad \; \; \, \mathbf{X}_{\text{trans}} \in \mathbb{R}^{(L+2) \times D} 
\end{flalign*}
}
where $x_v$ and $x_a$ are the valence and arousal inputs, $\mathbf{x}_t$ is the input token sequence, $\mathbf{b}$ represents the boundary offsets, and $\mathbf{W}_{\text{pe}}$ is the learned positional encoding. 
The feed-forward networks $\text{FFN}_v$, $\text{FFN}_a$, and $\text{FFN}_b$ process the valence, arousal, and boundary offset inputs, respectively. 
{\color{black}The plus sign $+$ denotes tensor addition. The first concatenation $[\,\cdot\,]_1$ is along the sequence (first) dimension producing a sequence of length $L+2$, and the second concatenation $[\,\cdot\,]_2$ is along the feature (second) dimension producing a feature of size $D$.}

Algorithm \ref{alg:offset} outlines the process of computing boundary offsets during inference, i.e., music generation. The model generates music autoregressively, producing one token per forward pass. We maintain a time cursor by tracking the generated \texttt{TIMESHIFT} tokens and compute the boundary offset, i.e., the time remaining until the next boundary, for each generated token. If the model generates a \texttt{CHORD} token, we calculate the absolute difference between the time cursor and each input boundary. The boundary with a difference smaller than the predefined sensitivity threshold $\xi$ is considered successfully generated. We then remove these boundaries from future offset calculations by replacing them with infinity. Next, we compute the boundary offset for the generated token, regardless of its type. 
This offset represents the distance to the next closest boundary. We compute it by subtracting the time cursor from each input boundary and replacing any negative values (corresponding to past boundaries) with infinity to ignore them. We then take the minimum of the resulting values as the offset to the nearest boundary and clamp it to a maximum value if necessary. The resulting boundary offset and generated token are appended to their respective lists to be used in the next timestep.
For simplicity, Algorithm \ref{alg:offset} is presented for a single sample, but in practice, this operation is performed in minibatches. During training, we preprocess the entire input sequence at once by constructing a time grid instead of a time cursor, allowing us to calculate boundary offsets for all tokens simultaneously. For clarity, the initial list of generated tokens is shown as empty; however, in practice, we begin with a \texttt{START} token.

The \texttt{ON} tokens that appear after a \texttt{CHORD} token and before the next \texttt{TIMESHIFT} token are considered notes of the generated chord. To make the generated chord more distinctive, we increase the velocity of these notes. Music generation continues until the time cursor reaches the duration of the video. 

\begin{algorithm}[t]
    \caption{Creating boundary offsets during music generation in inference.}
    \label{alg:offset}
    \textbf{Input}: List of input boundaries (in seconds), $\mathbf{b}$; video duration, $d$; valence, $x_v$; arousal $x_a$\\
    \textbf{Parameter}: Sensitivity, $\xi$; distances to all boundaries, $\boldsymbol{\delta}_b$; maximum offset, $\delta_{\max}$; generated boundary mask, $\mathbf{m_b}$; token type $t_t$; token value, $t_v$; time cursor, $c$; generation function including forward-pass and sampling, g\\
    \textbf{Output}: List of generated tokens $\mathbf{t}$; list of boundary offsets, $\boldsymbol{\delta}$
    \begin{algorithmic}
        \STATE Let $c=0$; $\boldsymbol{\delta}=\left[ \hspace{2pt} \right]$; $\mathbf{t}=\left[ \hspace{2pt} \right]$.
        \WHILE{$c < d$}
            \STATE \texttt{\# generate token as (type, value):} 
            \STATE $(t_t, t_v) = g(\mathbf{t}, \boldsymbol{\delta}, x_v, x_a)$
            \IF {$t_t$ == \texttt{TIMESHIFT}}
                \STATE $c = c + t_v $
            \ELSIF {$t_t$ == \texttt{CHORD}}
                \STATE $\mathbf{m_b} = \left| c - \mathbf{b} \right| < \xi$.
                \STATE $\mathbf{b}\left[\mathbf{m_b}\right] = +\infty$.
            \ENDIF

            \STATE $\boldsymbol{\delta}_b = \mathbf{b} - c $
            \STATE $\boldsymbol{\delta}_b[\boldsymbol{\delta}_b < 0] = +\infty $

            \STATE $\boldsymbol{\delta}$.append$(\min (\min (\boldsymbol{\delta}_b), \delta_{\max}))$
            
            \STATE $\mathbf{t}$.append$((t_t, t_v))$

        \ENDWHILE
    \end{algorithmic}
\end{algorithm}

\subsubsection{Scene detection}

During video-based inference, we first extract the video's temporal boundaries, specifically its scene cut locations. As scene detection is a {\color{black}well-studied problem with established solutions,} we employ the off-the-shelf FFmpeg software\footnote{https://www.ffmpeg.org/}. As we generate musical chords near scene boundaries, we aim to avoid an excessive number of densely spaced chords. We therefore apply a difference filter to the extracted scene cuts and remove those occurring less than $t_{\text{diff}}$ seconds apart. The remaining timestamps serve as the input boundaries for the music generator.

\subsubsection{Emotion mapping}

We use our pretrained emotion classifier~\cite{vemoclap} to extract emotions from input videos as probability distributions over Ekman's basic emotion categories~\cite{ekman}. 
Our music generator is conditioned on emotions represented as valence and arousal values ranging from -1 to 1, rather than discrete categories. Since the emotion classifier outputs discrete emotion categories, we first map them to the valence-arousal plane~\cite{valence_arousal} before feeding them into the music generator.

Russell and Mehrabian conducted user studies to find the corresponding valence and arousal values of discrete emotion categories~\cite{mapping}. They presented their findings in a table that contains the mean and standard deviation values of valence and arousal for each categorical emotion. We extracted and used the values corresponding to Ekman's six basic emotions~\cite{ekman}, which are presented in Table \ref{tab:mapping}. While Russell and Mehrabian presented one of the oldest user studies that examine the relationship between dimensional and categorical emotion representations, their study remains the most comprehensive in terms of number of users ($N=300$) and number of categorical emotions ($K=151$), and is still employed in recent machine learning applications~\cite{eeg}. Similar and newer user studies from Hoffmann et al. ($N=70$, $K=22$) \cite{hoffmann} and Trnka et al. ($N=187$, $K=16$) \cite{trnka} are more limited, and do not include all of Ekman's basic emotions.

Using the output probabilities of the video classifier along with the means and standard deviations of each emotion, we construct a mixture of Gaussian distributions. Valence-arousal values can be obtained either by sampling an emotion category based on the output probabilities and then drawing from its corresponding Gaussian, or by using the mixture mean computed as the weighted average of the category means. Additionally, we introduce a parameter $\mu_{\max}$ representing the maximum absolute value of means across all emotion categories. Using $\mu_{\max}$, we compute a single scaling coefficient applied uniformly to all valence-arousal distributions. This approach allows users to adjust the conditioning of the music generator, choosing between a wider or narrower range of emotions.

Although not used in our model, we also formalize a method to invert this mapping, that is, to map valence-arousal values to emotion categories by comparing the valence-arousal coordinates to each category’s distribution defined by its mean and standard deviation. This involves calculating a distance metric, such as Euclidean or Mahalanobis distance, or computing the likelihood assuming Gaussian distributions, and then assigning the category with the smallest distance or highest likelihood. This provides a generalizable framework for integrating dimensional and categorical emotion representations in multimodal systems.

\begin{table}[t]

\caption{Valence and arousal means and standard deviations for Ekman’s six basic emotions~\cite{ekman}, derived from Russell and Mehrabian’s user study~\cite{mapping}.}
\centering
\begin{tabular}{l | cc | cc}
 & \multicolumn{2}{c|}{\textbf{Valence}} & \multicolumn{2}{c}{\textbf{Arousal}} \\ 
\cline{2-3} \cline{4-5}
\textbf{Emotion} & \textbf{Mean} & \textbf{SD} & \textbf{Mean} & \textbf{SD} \\ 
\hline
anger     & -0.51           & 0.20 & \phantom{-}0.59 & 0.29 \\
disgust   & -0.60           & 0.20 & \phantom{-}0.35 & 0.41 \\
fear      & -0.64           & 0.20 & \phantom{-}0.60 & 0.32 \\
joy       & \phantom{-}0.76 & 0.22 & \phantom{-}0.48 & 0.26 \\
sadness   & -0.63           & 0.23 & -0.27           & 0.34 \\
surprise  & \phantom{-}0.40 & 0.30 & \phantom{-}0.67 & 0.27 \\
\end{tabular}

\label{tab:mapping}
\end{table}

\section{Experimental setup}

We perform objective and subjective evaluations of our overall model using two datasets, against open-source state-of-the-art models that generate MIDI from arbitrary videos. Comparison with V-MusProd~\cite{zhuo} and XMusic~\cite{xmusic} was not possible, as their code is not publicly available. We therefore compare our method against Video2Music~\cite{kang} and CMT~\cite{di}.

\subsection{Implementation details}

Our music generator consists of 11 layers, 8 attention heads, and a model dimensionality of 512. To ensure a fair comparison, we align our model size with the compared models: our model has 37M parameters, compared to 39M in CMT~\cite{di} and 33M in Video2Music~\cite{kang}. Training is done with a context length of 1216 and a batch size of 64. The model is implemented in PyTorch~\cite{pytorch} and trained on a single NVIDIA A100 80GB GPU. We train our model using cross-entropy loss with a learning rate of 2e-4 for the first 300k steps, followed by 5e-5 for the next 300k steps. We use the Adam optimizer with gradient clipping to a norm of 1~\cite{adam}. For data augmentation, we transpose the pitches of all instruments, except drums, by a randomly chosen number of semitones between -3 and 3, inclusive. To prevent the model from becoming overly reliant on the \texttt{CHORD} token for chord generation, we randomly remove 20\% of these tokens during training. Furthermore, we amplify the loss of the \texttt{CHORD} token by a factor of 10, as it influences multiple preceding chord notes and plays a critical role in structuring the generated music. We set the maximum boundary offset $\delta_{\max}$ to 8 seconds. We reserve 5\% of the training data for validation. These training hyperparameters are selected via grid search to minimize validation loss.

For video-based inference in evaluation, we set the maximum absolute value of the emotion category means $\mu_{\max}$ to 0.8. We obtain valence–arousal values as the mean of the weighted mixture of Gaussians derived from the predicted emotion categories. Although this conditioning is deterministic, the music generator remains stochastic due to probabilistic token sampling during generation. The difference filter discards scene timestamps that are less than $t_{\text{diff}}=4$ seconds apart. We set the boundary sensitivity threshold $\xi$ to 1 second. To ensure consistency, all methods generate MIDI outputs that are synthesized into audio waveforms using Fluidsynth\footnote{https://www.fluidsynth.com}, followed by peak normalization to -3 dB. We apply a 3-second fade-out effect to prevent an abrupt ending and ensure the music matches the duration of the input video. The audio is then merged with the input video by overwriting its audio channel. For all methods, outputs are generated in a single run across all datasets using a fixed random seed. The inference hyperparameters are chosen by manually inspecting outputs on videos not included in the evaluation.

\subsection{Evaluation datasets}

For objective and subjective evaluation, we first use the \mbox{EmoMV-C} dataset, which consists of music videos containing music audio tracks to be used as ground truth for objective evaluation~\cite{emomv}. We exclude auditory features during the emotion classification to prevent data leakage. To assess the model's generalization across different video types, we also include the Pittsburgh Advertisements (Ads) dataset, which contains advertisement videos that may or may not feature music~\cite{ads}. As advertisements often rely on music to maximize viewer engagement, the Ads dataset provides a real-world scenario for evaluation~\cite{ads_music}.

We use the full EmoMV-C validation split, which contains 48 thirty-second videos. We filter the Ads dataset to match the total duration of EmoMV-C. We select videos associated with the four basic emotions commonly used in music emotion classification~\cite{music_emotion}, covering the quadrants of Russell's valence-arousal model~\cite{valence_arousal}: \textit{cheerful}, \textit{calm}, \textit{angry}, and \textit{sad}, providing a broad coverage of the emotional space for evaluation. While some of the chosen emotions do overlap with Ekman's categories~\cite{ekman}, we avoid explicitly using Ekman's emotions to select the evaluation videos. This prevents biasing the evaluation toward our method, since our video emotion classifier is trained to output Ekman's categories. We exclude videos shorter than 1 minute. To ensure an unbiased selection, we use YouTube IDs, which are generated randomly by YouTube. We sort these IDs alphabetically and select the first six videos from each emotion category, resulting in 24 evaluation videos. Finally, we trim the videos to a uniform duration of 1 minute. The resulting evaluation videos are then fed into the compared models.

\begin{figure*}[t] 
    \centering
    \includegraphics[width=0.99\textwidth]{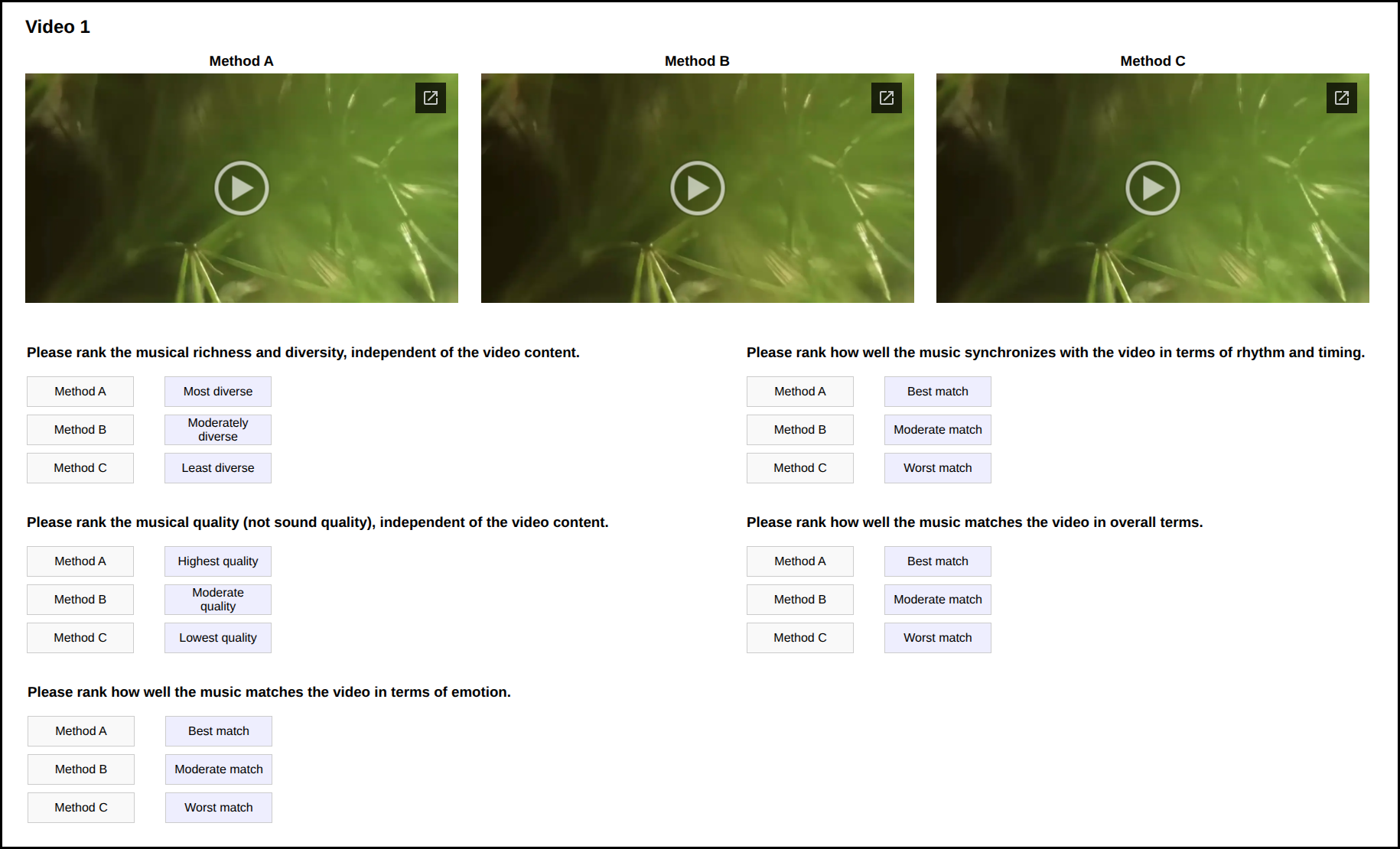}

    \caption{Sample view of the survey used in the subjective evaluation. Participants were asked to drag the method boxes from the left onto the ranking boxes on the right.}
    \label{fig:survey}

\end{figure*}

\subsection{Objective evaluation}

We first evaluate music similarity using the ground-truth audio from the EmoMV-C dataset. Although the ground truth is not in MIDI, all methods generate MIDI outputs that are synthesized to audio, enabling a fair comparison in the audio domain. We evaluate the Kullback-Leibler Divergence (KL) between the labels of generated and ground-truth music, calculated using a pretrained music tagging model~\cite{harmoniccnn}. This measures how closely the distribution of semantic tags in the generated music matches that of the ground-truth. We avoid the commonly used Fréchet Audio Distance, as it compares overall audio distributions suited for unconditional generation~\cite{fad}, whereas video-based music generation requires per-sample evaluation to assess alignment with each video's unique content.

We additionally evaluate temporal audio-video alignment (AV-Align) score, which is the Intersection-over-Union between video and audio peaks~\cite{avalign}, using both EmoMV-C and Ads datasets. Video peaks are detected from the mean optical flow magnitude~\cite{flow}, while audio peaks correspond to audio onsets~\cite{onsets}. We calculate AV-Align scores at two temporal resolutions (windows): one video frame (33.3 ms at 30 FPS) and one second. For each window, local peaks in both video and audio are detected and matched within the same temporal range to assess alignment. {\color{black}As an internal evaluation, we additionally report the alignment between the generated MIDI chords (specifically the \texttt{CHORD} token) and video scene cuts using the same 33.3~ms and 1~s windows. As baseline methods lack an explicit boundary-matching architecture, these alignment scores are reported exclusively for EMSYNC to validate the effectiveness of its temporal conditioning mechanism.} Finally, we report the average runtime for each model when generating music for a one-minute video, including model initialization, {\color{black}video analysis, and music generation}.

\subsection{Subjective evaluation}

Using the evaluation videos from the EmoMV-C and the Ads datasets, we generate accompanying music with the three models and create survey pages, each containing two videos. For each video, the three music versions generated by the compared models are presented side by side with anonymized model names, with each model's output appearing an equal number of times in the left, center, and right positions. We also provide output samples from eight videos online, with model names deanonymized and each model's output appearing in the same position for clarity.\footnote{https://serkansulun.com/emsync}

We enrolled 153 remote participants who provided informed consent before taking part in the survey. They were asked to rank the three models based on the standard criteria used in previous works~\cite{di,kang}:

\begin{itemize}
    \item \textit{Music Richness (MR)} — the richness and diversity of the music, independent of the video content.
    \item \textit{Music Quality (MQ)} — the overall quality of the music, independent of the video content.
    \item \textit{Emotion Match (EM)} — how well the music matches the video in terms of emotion.
    \item \textit{Timing Match (TM)} — how well the music synchronizes with the video in terms of rhythm and timing.
    \item \textit{Overall Match (OM)} — how well the music matches the video as a whole.
\end{itemize}

 Each model is assigned a unique ranking of 1 (best), 2, or 3 (worst). Figure \ref{fig:survey} shows a sample view of the user survey.

\section{Results and conclusion}

\begin{table}[t]
\caption{Objective comparison with the state-of-the-art}
\centering
\begin{tabular}{l||ccc|cc|c}
Dataset & \multicolumn{3}{c|}{EmoMV-C} & \multicolumn{2}{c|}{Ads} &  \\
\hline
Metric & AV$_f$ & AV$_s$ & KL & AV$_f$ & AV$_s$ & RT \\
\hline
CMT           & 0.014 & 0.150 & 4.728 & 0.008 & 0.099 & 3.55 \\
Video2Music   & 0.010 & \textbf{0.245} & 9.789 & 0.002 & 0.107 & 1.61 \\
EMSYNC (Ours) & \textbf{0.017} & 0.233 & \textbf{4.175} & \textbf{0.011} & \textbf{0.150} & \textbf{1.42} \\
\end{tabular}
\label{tab:objective}
\end{table}

Table \ref{tab:objective} presents objective evaluation results for EMSYNC, Video2Music~\cite{kang}, and CMT~\cite{di}, on the EmoMV-C and Ads datasets. AV$_f$ and AV$_s$ denote audio-video alignment in frame and second-levels, respectively, where higher is better. KL denotes the KL-divergence, where lower is better. RT denotes average runtime in minutes to generate music for a 1-minute video, {\color{black}including model initialization, video analysis, and music generation,} where lower is better. EMSYNC outperforms both baselines across a majority of the metrics. On EmoMV-C, it achieves the best frame-level alignment and the lowest KL divergence, indicating superior synchronization to video and similarity to ground-truth music. While Video2Music performs best on second-level alignment, EMSYNC remains competitive. On the Ads dataset, EMSYNC obtains the best scores across both alignment metrics, further demonstrating its robustness across different video domains. We don't calculate KL divergence because the Ads dataset doesn't include ground truth music. EMSYNC also achieves the fastest runtime among the evaluated methods.
{\color{black}For the EmoMV-C and Ads datasets, EMSYNC achieves chord-scene cut alignment scores of 0.297 and 0.317 at the 33.3~ms resolution, and 0.899 and 0.909 within a one-second window, respectively, demonstrating its capabilities in aligning temporal boundaries through its temporal conditioning mechanism.}

\begin{table}[t]
\caption{Subjective comparison with the state-of-the-art:\\ Mean ranks ± 95\% confidence margins}
\centering
\begin{threeparttable}
\begin{tabular}{c|c|ccc}
Dataset & Metric & CMT & Video2Music & \makecell{EMSYNC \\ (Ours)} \\
\hline \hline
\multirow{5}{*}{\rotatebox[origin=c]{90}{EmoMV-C}}
& MR  & 1.870 ± 0.102 & 2.317 ± 0.108 & \textbf{1.812} ± 0.112 \\
& MQ  & 2.226 ± 0.099 & 1.952 ± 0.116 & \textbf{1.822} ± 0.113 \\
& EM  & 2.111 ± 0.100 & 2.043 ± 0.115 & \textbf{1.846} ± 0.117 \\
& TM  & 2.058 ± 0.098 & 2.067 ± 0.115 & \textbf{1.875} ± 0.120 \\
& OM  & 2.106 ± 0.100 & 2.087 ± 0.114 & \textbf{1.808} ± 0.116 \\
\hline
\multirow{5}{*}{\rotatebox[origin=c]{90}{Ads}}
& MR  & 1.949 ± 0.153 & 2.439 ± 0.155 & \textbf{1.612} ± 0.140 \\
& MQ  & 2.469 ± 0.150 & 1.888 ± 0.144 & \textbf{1.643} ± 0.153 \\
& EM  & 2.214 ± 0.153 & 2.051 ± 0.171 & \textbf{1.735} ± 0.154 \\
& TM  & 2.245 ± 0.168 & 2.010 ± 0.159 & \textbf{1.745} ± 0.150 \\
& OM  & 2.255 ± 0.153 & 2.122 ± 0.166 & \textbf{1.622} ± 0.145 \\
\end{tabular}
\label{tab:subjective}
\end{threeparttable}
\end{table}

Table \ref{tab:subjective} demonstrates the subjective results in terms of average user rankings with 95\% confidence margins. A lower average score represents a better ranking, with 1 being the best and 3 the worst possible. EMSYNC achieves the best average ranking across all metrics and datasets. The difference is more pronounced in video-music correspondence metrics such as emotion match and timing match, highlighting our model’s stronger alignment to video. For the overall match, our method outperforms the others with non-overlapping 95\% confidence intervals, reflecting a decisive user preference for EMSYNC in video-based music generation. 

\begin{figure}[t] 
    \centering
    \includegraphics[width=0.99\columnwidth]{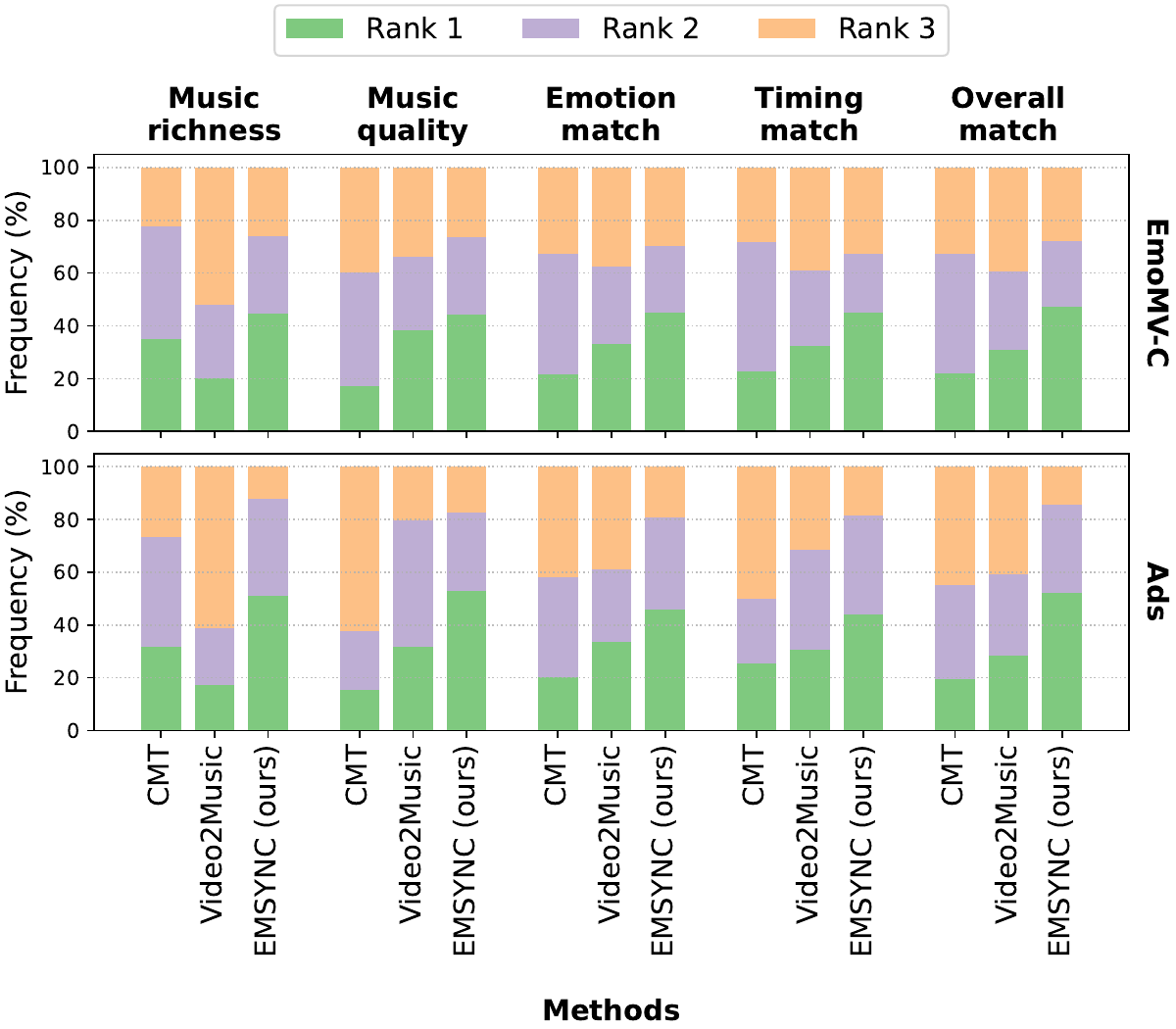}
    \caption{Distributions of rankings assigned by the participants in the subjective evaluation using EmoMV-C (top) and Ads (bottom) datasets.}
    \label{fig:bar}
\end{figure}

Figure~\ref{fig:bar} further shows the distribution of rankings assigned by participants in the subjective evaluation. EMSYNC achieves the highest frequency of the top rank (Rank 1) across all metrics and datasets.

These results highlight EMSYNC’s effectiveness in generating music that aligns with the input video both temporally and emotionally, across objective and subjective evaluations. By automatically generating soundtracks for user-provided videos, our work has the potential to streamline video production and enhance viewer engagement while offering a valuable framework for both machine learning researchers and multimedia content creators. In future work, we aim to incorporate more complex musical boundaries, such as those that separate melodic segments {\color{black}or motifs}. We will also enable dynamic emotional conditioning to adapt the music continuously to evolving complex emotional content in videos.

\section{Acknowledgments}
\noindent This work is co-financed by Component 5 - Capitalization and Business Innovation, integrated in the Resilience Dimension of the Recovery and Resilience Plan within the scope of the Recovery and Resilience Mechanism (MRR) of the European Union (EU), framed in the Next Generation EU, for the period 2021 - 2026, within project NEXUS, with reference 53, and the fellowship from FCT - Fundação para a Ciência e a Tecnologia (2022.09594.BD).

\bibliographystyle{IEEEtran}
\bibliography{references}

@inproceedings{transformer,
  author       = {Ashish Vaswani and
                  Noam Shazeer and
                  Niki Parmar and
                  Jakob Uszkoreit and
                  Llion Jones and
                  Aidan N. Gomez and
                  Lukasz Kaiser and
                  Illia Polosukhin},
  title        = {Attention is All you Need},
  booktitle    = {Advances in Neural Information Processing Systems},
  pages        = {5998--6008},
  year         = {2017}
}

@inproceedings{vemoclap,
  author={Sulun, Serkan and Viana, Paula and Davies, Matthew E. P.},
  booktitle={2024 International Symposium on Multimedia (ISM)}, 
  title={{VEMOCLAP}: A video emotion classification web application}, 
  year={2024},
  volume={},
  number={},
  pages={137-140},
  doi={10.1109/ISM63611.2024.00029}
}

@article{trailer,
  title = {Movie Trailer Genre Classification Using Multimodal Pretrained Features},
  author = {Sulun, Serkan and Viana, Paula and Davies, Matthew E. P.},
  year = {2024},
  journal = {Expert Systems with Applications},
  volume = {258},
  pages = {125209},
  issn = {0957-4174},
  doi = {10.1016/j.eswa.2024.125209}
}

@inproceedings{clip,
  title={Learning transferable visual models from natural language supervision},
  author={Radford, Alec and Kim, Jong Wook and Hallacy, Chris and Ramesh, Aditya and Goh, Gabriel and Agarwal, Sandhini and Sastry, Girish and Askell, Amanda and Mishkin, Pamela and Clark, Jack and others},
  booktitle={Proceedings of the 38th International Conference on Machine Learning, ICML 2021},
  pages={8748--8763},
  year={2021},
  organization={PMLR}
}

@article{ekman6,
  title={Heterogeneous knowledge transfer in video emotion recognition, attribution and summarization},
  author={Xu, Baohan and Fu, Yanwei and Jiang, Yu-Gang and Li, Boyang and Sigal, Leonid},
  journal={IEEE Transactions on Affective Computing},
  volume={9},
  number={2},
  pages={255--270},
  year={2016},
  publisher={IEEE}
}

@article{valence_arousal,
  title = {A Circumplex Model of Affect.},
  author = {Russell, James A.},
  year = {1980},
  journal = {Journal of personality and social psychology},
  volume = {39},
  number = {6},
  pages = {1161},
  publisher = {American Psychological Association}
}

@article{mapping,
  title = {Evidence for a Three-Factor Theory of Emotions},
  author = {Russell, James A. and Mehrabian, Albert},
  year = {1977},
  journal = {Journal of research in Personality},
  volume = {11},
  number = {3},
  pages = {273--294},
  publisher = {Elsevier}
}

@article{ekman,
  title = {Universals and Cultural Differences in Facial Expressions of Emotion},
  author = {Ekman, Paul},
  year = {1971},
  journal = {Nebraska Symposium on Motivation},
  volume = {19},
  pages = {207--283},
  publisher = {University of Nebraska Press},
  address = {US},
  issn = {0146-7875}
}

@inproceedings{musictransformer,
  author       = {Cheng{-}Zhi Anna Huang and
                  Ashish Vaswani and
                  Jakob Uszkoreit and
                  Ian Simon and
                  Curtis Hawthorne and
                  Noam Shazeer and
                  Andrew M. Dai and
                  Matthew D. Hoffman and
                  Monica Dinculescu and
                  Douglas Eck},
  title        = {Music Transformer: Generating Music with Long-Term Structure},
  booktitle    = {7th International Conference on Learning Representations, {ICLR} 2019},
  publisher    = {OpenReview.net},
  year         = {2019},
}

@article{event_encoding,
  title = {This Time with Feeling: Learning Expressive Musical Performance},
  author = {Oore, Sageev and Simon, Ian and Dieleman, Sander and Eck, Douglas and Simonyan, Karen},
  year = {2020},
  journal = {Neural Computing and Applications},
  volume = {32},
  number = {4},
  pages = {955--967},
  publisher = {Springer}
}

@inproceedings{lpd,
  title={{Musegan}: Multi-track sequential generative adversarial networks for symbolic music generation and accompaniment},
  author={Dong, Hao-Wen and Hsiao, Wen-Yi and Yang, Li-Chia and Yang, Yi-Hsuan},
  booktitle={Proceedings of the AAAI Conference on Artificial Intelligence},
  volume={32},
  year={2018}
}

@phdthesis{lmd,
  title = {Learning-Based Methods for Comparing Sequences, with Applications to Audio-to-Midi Alignment and Matching},
  author = {Raffel, Colin},
  year = {2016},
  school = {Columbia University, New York, NY, USA},
  type = {{Ph.D. dissertation}}
}

@article{access,
  title = {Symbolic Music Generation Conditioned on Continuous-Valued Emotions},
  author = {Sulun, Serkan and Davies, Matthew E. P. and Viana, Paula},
  year = {2022},
  journal = {IEEE Access},
  volume = {10},
  pages = {44617--44626},
  doi = {10.1109/ACCESS.2022.3169744}
}

@inproceedings{learned_position,
  author       = {Benyou Wang and
                  Donghao Zhao and
                  Christina Lioma and
                  Qiuchi Li and
                  Peng Zhang and
                  Jakob Grue Simonsen},
  title        = {Encoding word order in complex embeddings},
  booktitle    = {8th International Conference on Learning Representations, ICLR 2020},
  year         = {2020},
  bibsource    = {dblp computer science bibliography, https://dblp.org}
}

@inproceedings{di,
author = {Di, Shangzhe and Jiang, Zeren and Liu, Si and Wang, Zhaokai and Zhu, Leyan and He, Zexin and Liu, Hongming and Yan, Shuicheng},
title = {Video Background Music Generation with Controllable Music Transformer},
year = {2021},
isbn = {9781450386517},
publisher = {Association for Computing Machinery},
doi = {10.1145/3474085.3475195},
booktitle = {Proceedings of the 29th ACM International Conference on Multimedia},
pages = {2037–2045},
series = {MM '21}
}

@inproceedings{foley,
  title = {Foley Music: Learning to Generate Music from Videos},
  shorttitle = {Foley Music},
  booktitle = {Computer Vision - ECCV 2020 - 16th European Conference},
  author = {Gan, Chuang and Huang, Deng and Chen, Peihao and Tenenbaum, Joshua B. and Torralba, Antonio},
  year = {2020},
  volume = {12356},
  pages = {758--775},
  publisher = {Springer},
  doi = {10.1007/978-3-030-58621-8_44}
}

@inproceedings{sighttosound,
  title = {Sight to Sound: An End-to-End Approach for Visual Piano Transcription},
  shorttitle = {Sight to Sound},
  booktitle = {IEEE International Conference on Acoustics, Speech and Signal Processing, ICASSP 2020},
  author = {Koepke, A. Sophia and Wiles, Olivia and Moses, Yael and Zisserman, Andrew},
  year = {2020},
  pages = {1838--1842},
  publisher = {IEEE},
  doi = {10.1109/ICASSP40776.2020.9053115}
}

@inproceedings{audeo,
  title = {Audeo: Audio Generation for a Silent Performance Video},
  shorttitle = {Audeo},
  booktitle = {Advances in Neural Information Processing Systems},
  author = {Su, Kun and Liu, Xiulong and Shlizerman, Eli},
  year = {2020},
  volume = {33},
  pages = {3325--3337}
}

@inproceedings{rhythmicnet,
 author = {Su, Kun and Liu, Xiulong and Shlizerman, Eli},
 booktitle = {Advances in Neural Information Processing Systems},
 pages = {29258--29273},
 publisher = {Curran Associates, Inc.},
 title = {How Does it Sound?},
 volume = {34},
 year = {2021}
}

@inproceedings{zhuo,
  title = {Video Background Music Generation: Dataset, Method and Evaluation},
  shorttitle = {Video Background Music Generation},
  booktitle = {{IEEE/CVF} International Conference on Computer Vision, ICCV 2023},
  author = {Zhuo, Le and Wang, Zhaokai and Wang, Baisen and Liao, Yue and Bao, Chenxi and Peng, Stanley and Han, Songhao and Zhang, Aixi and Fang, Fei and Liu, Si},
  year = {2023},
  pages = {15591--15601},
  publisher = {{IEEE}},
  doi = {10.1109/ICCV51070.2023.01433}
}

@article{kang,
  title = {Video2Music: Suitable Music Generation from Videos Using an Affective Multimodal Transformer Model},
  shorttitle = {Video2Music},
  author = {Kang, Jaeyong and Poria, Soujanya and Herremans, Dorien},
  year = {2024},
  month = sep,
  journal = {Expert Systems with Applications},
  volume = {249},
  pages = {123640},
  issn = {0957-4174},
  doi = {10.1016/j.eswa.2024.123640},
  urldate = {2024-04-10}
}

@inproceedings{music_dataset,
  title = {The Sound of Pixels},
  booktitle = {Computer Vision - ECCV 2018 - 14th European Conference},
  author = {Zhao, Hang and Gan, Chuang and Rouditchenko, Andrew and Vondrick, Carl and McDermott, Josh and Torralba, Antonio},
  year = {2018},
  pages = {570--586}
}

@article{video_midi_dataset,
  title = {Creating a Multitrack Classical Music Performance Dataset for Multimodal Music Analysis: Challenges, Insights, and Applications},
  shorttitle = {Creating a Multitrack Classical Music Performance Dataset for Multimodal Music Analysis},
  author = {Li, Bochen and Liu, Xinzhao and Dinesh, Karthik and Duan, Zhiyao and Sharma, Gaurav},
  year = {2018},
  journal = {IEEE Transactions on Multimedia},
  volume = {21},
  number = {2},
  pages = {522--535},
  publisher = {IEEE}
}

@inproceedings{gcn,
  title = {Semi-Supervised Classification with Graph Convolutional Networks},
  booktitle = {5th International Conference on Learning Representations, ICLR 2017},
  author = {Kipf, Thomas N. and Welling, Max},
  year = {2017},
}

@inproceedings{compound_words,
  title = {Compound Word Transformer: Learning to Compose Full-Song Music over Dynamic Directed Hypergraphs},
  shorttitle = {Compound Word Transformer},
  booktitle = {Proceedings of the AAAI Conference on Artificial Intelligence},
  author = {Hsiao, Wen-Yi and Liu, Jen-Yu and Yeh, Yin-Cheng and Yang, Yi-Hsuan},
  year = {2021},
  pages = {178--186},
  doi = {10.1609/AAAI.V35I1.16091},
  urldate = {2025-01-24}
}

@inproceedings{onset_and_frames,
  title = {Onsets and Frames: Dual-Objective Piano Transcription},
  shorttitle = {Onsets and Frames},
  booktitle = {Proceedings of the 19th International Society for Music Information Retrieval Conference},
  author = {Hawthorne, Curtis and Elsen, Erich and Song, Jialin and Roberts, Adam and Simon, Ian and Raffel, Colin and Engel, Jesse H. and Oore, Sageev and Eck, Douglas},
  year = {2018},
  pages = {50--57}
}

@inproceedings{musicgen1,
author = {Lam, Max W. Y. and Tian, Qiao and Li, Tang and Yin, Zongyu and Feng, Siyuan and Tu, Ming and Ji, Yuliang and Xia, Rui and Ma, Mingbo and Song, Xuchen and Chen, Jitong and Wang, Yuping and Wang, Yuxuan},
title = {Efficient neural music generation},
year = {2023},
publisher = {Curran Associates Inc.},
booktitle = {Proceedings of the 37th International Conference on Neural Information Processing Systems},
articleno = {766},
numpages = {14},
series = {NIPS '23}
}

@inproceedings{musicgen2,
author = {Copet, Jade and Kreuk, Felix and Gat, Itai and Remez, Tal and Kant, David and Synnaeve, Gabriel and Adi, Yossi and D\'{e}fossez, Alexandre},
title = {Simple and controllable music generation},
year = {2023},
publisher = {Curran Associates Inc.},
booktitle = {Proceedings of the 37th International Conference on Neural Information Processing Systems},
articleno = {2066},
numpages = {17},
series = {NIPS '23}
}

@inproceedings{swin,
  title = {Video Swin Transformer},
  booktitle = {IEEE/CVF Conference on Computer Vision and Pattern Recognition, CVPR 2022},
  author = {Liu, Ze and Ning, Jia and Cao, Yue and Wei, Yixuan and Zhang, Zheng and Lin, Stephen and Hu, Han},
  year = {2022},
  pages = {3192--3201},
  publisher = {IEEE},
  doi = {10.1109/CVPR52688.2022.00320}
}

@inproceedings{vivit,
  title={Vivit: A video vision transformer},
  author={Arnab, Anurag and Dehghani, Mostafa and Heigold, Georg and Sun, Chen and Lu{\v{c}}i{\'c}, Mario and Schmid, Cordelia},
  booktitle={IEEE/CVF international conference on computer vision, CVPR 2021},
  pages={6836--6846},
  year={2021}
}

@inproceedings{midiemotion1,
  title = {Multi-Modal Music Emotion Recognition: A New Dataset, Methodology and Comparative Analysis},
  shorttitle = {Multi-Modal Music Emotion Recognition},
  booktitle = {International Symposium on Computer Music Multidisciplinary Research},
  author = {Panda, Renato and Malheiro, Ricardo and Rocha, Bruno and Oliveira, Ant{\'o}nio and Paiva, Rui Pedro},
  year = {2013},
}

@inproceedings{midiemotion3,
  title = {{EMOPIA}: A Multi-Modal Pop Piano Dataset For Emotion Recognition and Emotion-Based Music Generation},
  shorttitle = {EMOPIA},
  booktitle = {Proceedings of the 22nd International Society for Music Information Retrieval Conference, ISMIR 2021},
  author = {Hung, Hsiao-Tzu and Ching, Joann and Doh, Seungheon and Kim, Nabin and Nam, Juhan and Yang, Yi-Hsuan},
  year = {2021},
  pages = {318--325},
  urldate = {2021-12-14}
}

@article{chords,
  title = {The Representation of Harmonic Structure in Music: Hierarchies of Stability as a Function of Context},
  shorttitle = {The Representation of Harmonic Structure in Music},
  author = {Bharucha, Jamshed and Krumhansl, Carol L.},
  year = {1983},
  journal = {Cognition},
  volume = {13},
  number = {1},
  pages = {63--102},
  publisher = {Elsevier}
}

@inproceedings{position1,
  title = {Transformer Language Models without Positional Encodings Still Learn Positional Information},
  booktitle = {Findings of the Association for Computational Linguistics: EMNLP 2022},
  author = {Haviv, Adi and Ram, Ori and Press, Ofir and Izsak, Peter and Levy, Omer},
  year = {2022},
  pages = {1382--1390},
  publisher = {Association for Computational Linguistics},
  doi = {10.18653/V1/2022.FINDINGS-EMNLP.99},
  urldate = {2025-01-31}
}

@inproceedings{position2,
 author = {Kazemnejad, Amirhossein and Padhi, Inkit and Natesan Ramamurthy, Karthikeyan and Das, Payel and Reddy, Siva},
 booktitle = {Advances in Neural Information Processing Systems},
 pages = {24892--24928},
 publisher = {Curran Associates, Inc.},
 title = {The Impact of Positional Encoding on Length Generalization in Transformers},
 volume = {36},
 year = {2023}
}

@inproceedings{ads,
  title = {Automatic Understanding of Image and Video Advertisements},
  booktitle = {IEEE/CVF Conference on Computer Vision and Pattern Recognition, {CVPR} 2017},
  author = {Hussain, Zaeem and Zhang, Mingda and Zhang, Xiaozhong and Ye, Keren and Thomas, Christopher and Agha, Zuha and Ong, Nathan and Kovashka, Adriana},
  year = {2017},
  pages = {1705--1715}
}

@article{ads_music,
  title = {Musical Influences in Advertising: How Music Modifies First Impressions of Product Endorsers and Brands},
  shorttitle = {Musical Influences in Advertising},
  author = {Zander, Mark F.},
  year = {2006},
  month = oct,
  journal = {Psychology of Music},
  volume = {34},
  number = {4},
  pages = {465--480},
  issn = {0305-7356, 1741-3087},
  doi = {10.1177/0305735606067158},
  urldate = {2025-02-01},
  copyright = {https://journals.sagepub.com/page/policies/text-and-data-mining-license},
  langid = {english}
}

@incollection{music_emotion,
  author       = {Yi{-}Hsuan Yang and
                  Yu{-}Ching Lin and
                  Heng Tze Cheng and
                  I{-}Bin Liao and
                  Yeh{-}Chin Ho and
                  Homer H. Chen},
  title        = {Toward Multi-modal Music Emotion Classification},
  booktitle    = {Advances in Multimedia Information Processing - {PCM} 2008},
  series       = {Lecture Notes in Computer Science},
  volume       = {5353},
  pages        = {70--79},
  publisher    = {Springer},
  year         = {2008},
  doi          = {10.1007/978-3-540-89796-5\_8}
}

@inproceedings{adam,
  author       = {Diederik P. Kingma and
                  Jimmy Ba},
  title        = {Adam: {A} Method for Stochastic Optimization},
  booktitle    = {3rd International Conference on Learning Representations, {ICLR} 2015},
  year         = {2015}
}

@article{multimedia,
  title = {Current Trends in Consumption of Multimedia Content Using Online Streaming Platforms: A User-Centric Survey},
  shorttitle = {Current Trends in Consumption of Multimedia Content Using Online Streaming Platforms},
  author = {{Falkowski-Gilski}, Przemys{\l}aw and Uhl, Tadeus},
  year = {2020},
  journal = {Computer Science Review},
  volume = {37},
  pages = {100268},
  issn = {1574-0137},
  doi = {10.1016/j.cosrev.2020.100268},
  urldate = {2025-02-10}
}

@book{soundtrack,
  title={Theories of the Soundtrack},
  author={Buhler, James},
  year={2018},
  publisher={Oxford University Press}
}

@inproceedings{avalign,
  title = {Diverse and Aligned Audio-to-Video Generation via Text-to-Video Model Adaptation},
  booktitle={Proceedings of the AAAI Conference on Artificial Intelligence},
  author = {Yariv, Guy and Gat, Itai and Benaim, Sagie and Wolf, Lior and Schwartz, Idan and Adi, Yossi},
  year = {2024},
  pages = {6639--6647},
  doi = {10.1609/AAAI.V38I7.28486},
  urldate = {2025-05-21}
}

@article{emomv,
  title = {EmoMV: Affective Music-Video Correspondence Learning Datasets for Classification and Retrieval},
  shorttitle = {EmoMV},
  author = {Thao, Ha Thi Phuong and Roig, Gemma and Herremans, Dorien},
  year = {2023},
  month = mar,
  journal = {Information Fusion},
  volume = {91},
  pages = {64--79},
  issn = {1566-2535},
  doi = {10.1016/j.inffus.2022.10.002},
  urldate = {2025-05-21}
}

@inproceedings{harmoniccnn,
  title = {Data-Driven Harmonic Filters for Audio Representation Learning},
  booktitle = {ICASSP 2020 - 2020 IEEE International Conference on Acoustics, Speech and Signal Processing (ICASSP)},
  author = {Won, Minz and Chun, Sanghyuk and Nieto, Oriol and Serrc, Xavier},
  year = {2020},
  month = may,
  pages = {536--540},
  issn = {2379-190X},
  doi = {10.1109/ICASSP40776.2020.9053669},
  urldate = {2025-07-16}
}

@inproceedings{fad,
  title = {Fr{\'e}chet Audio Distance: A Reference-Free Metric for Evaluating Music Enhancement Algorithms.},
  shorttitle = {Fr{\'e}chet Audio Distance},
  booktitle = {INTERSPEECH},
  author = {Kilgour, Kevin and Zuluaga, Mauricio and Roblek, Dominik and Sharifi, Matthew},
  year = {2019},
  pages = {2350--2354}
}

@article{flow,
  title={Determining optical flow},
  author={Horn, Berthold KP and Schunck, Brian G},
  journal={Artificial intelligence},
  volume={17},
  number={1-3},
  pages={185--203},
  year={1981},
  publisher={Elsevier}
}

@inproceedings{onsets,
  title={Maximum filter vibrato suppression for onset detection},
  author={B{\"o}ck, Sebastian and Widmer, Gerhard},
  booktitle={16th International Conference on Digital Audio Effects (DAFx)},
  volume={7},
  pages={4},
  year={2013},
  organization={Citeseer}
}

@article{emotion_motivation1,
  title = {The Soundtrack (Putting Music in Its Place)},
  author = {Deutsch, Stephen},
  year = {2007},
  journal = {The Soundtrack},
  volume = {1},
  number = {1},
  pages = {3--13},
  urldate = {2025-04-15}
}

@article{temporal_motivation1,
  title = {Understanding Musical Soundtracks},
  author = {Cohen, Annabel J.},
  year = {1990},
  month = jul,
  journal = {Empirical Studies of the Arts},
  volume = {8},
  number = {2},
  pages = {111--124},
  issn = {0276-2374, 1541-4493},
  doi = {10.2190/8Y6G-KTM8-VDX4-UHRW},
  urldate = {2025-04-15},
  copyright = {https://journals.sagepub.com/page/policies/text-and-data-mining-license},
  langid = {english}
}

@book{temporal_motivation2,
  title = {Film Music: A Neglected Art: A Critical Study of Music in Films},
  shorttitle = {Film Music},
  author = {Prendergast, Roy M.},
  year = {1992},
  publisher = {WW Norton \& Company}
}

@inproceedings{pytorch,
 author = {Paszke, Adam and Gross, Sam and Massa, Francisco and Lerer, Adam and Bradbury, James and Chanan, Gregory and Killeen, Trevor and Lin, Zeming and Gimelshein, Natalia and Antiga, Luca and Desmaison, Alban and Kopf, Andreas and Yang, Edward and DeVito, Zachary and Raison, Martin and Tejani, Alykhan and Chilamkurthy, Sasank and Steiner, Benoit and Fang, Lu and Bai, Junjie and Chintala, Soumith},
 booktitle = {Advances in Neural Information Processing Systems},
 pages = {},
 publisher = {Curran Associates, Inc.},
 title = {PyTorch: An Imperative Style, High-Performance Deep Learning Library},
 volume = {32},
 year = {2019}
}

@incollection{eeg,
author = {Krzysztof Kotowski and Katarzyna Stapor},
title = {Machine Learning and EEG for Emotional State Estimation},
booktitle = {The Science of Emotional Intelligence},
publisher = {IntechOpen},
address = {London},
year = {2021},
chapter = {6},
doi = {10.5772/intechopen.97133}
}

@inproceedings{hoffmann,
  title = {Mapping Discrete Emotions into the Dimensional Space: An Empirical Approach},
  shorttitle = {Mapping Discrete Emotions into the Dimensional Space},
  booktitle = {2012 IEEE International Conference on Systems, Man, and Cybernetics (SMC)},
  author = {Hoffmann, Holger and Scheck, Andreas and Schuster, Timo and Walter, Steffen and Limbrecht, Kerstin and Traue, Harald C. and Kessler, Henrik},
  year = {2012},
  month = oct,
  pages = {3316--3320},
  issn = {1062-922X},
  doi = {10.1109/ICSMC.2012.6378303},
  urldate = {2025-10-02}
}

@article{trnka,
  title = {Modeling Semantic Emotion Space Using a 3D Hypercube-Projection: An Innovative Analytical Approach for the Psychology of Emotions},
  shorttitle = {Modeling Semantic Emotion Space Using a 3D Hypercube-Projection},
  author = {Trnka, Radek and La{\v c}ev, Alek and Balcar, Karel and Ku{\v s}ka, Martin and Tavel, Peter},
  year = {2016},
  journal = {Frontiers in psychology},
  volume = {7},
  pages = {522},
  publisher = {Frontiers Media SA}
}

@article{xmusic,
  title = {XMusic: Towards a Generalized and Controllable Symbolic Music Generation Framework},
  shorttitle = {XMusic},
  author = {Tian, Sida and Zhang, Can and Yuan, Wei and Tan, Wei and Zhu, Wenjie},
  year = {2025},
  journal = {IEEE Transactions on Multimedia},
  volume = {27},
  pages = {6857--6871},
  issn = {1941-0077},
  doi = {10.1109/TMM.2025.3590912},
  urldate = {2025-10-14}
}

@inproceedings{epia,
  title = {{Emotion4MIDI}: A Lyrics-Based Emotion-Labeled Symbolic Music Dataset},
  booktitle = {Progress in Artificial Intelligence},
  author = {Sulun, Serkan and Oliveira, Pedro and Viana, Paula},
  year = {2023},
  pages = {77--89},
  publisher = {Springer Nature Switzerland},
  address = {Cham},
  isbn = {978-3-031-49011-8}
}

\begin{IEEEbiography}[{\includegraphics[width=1in,height=1.25in,clip,keepaspectratio]{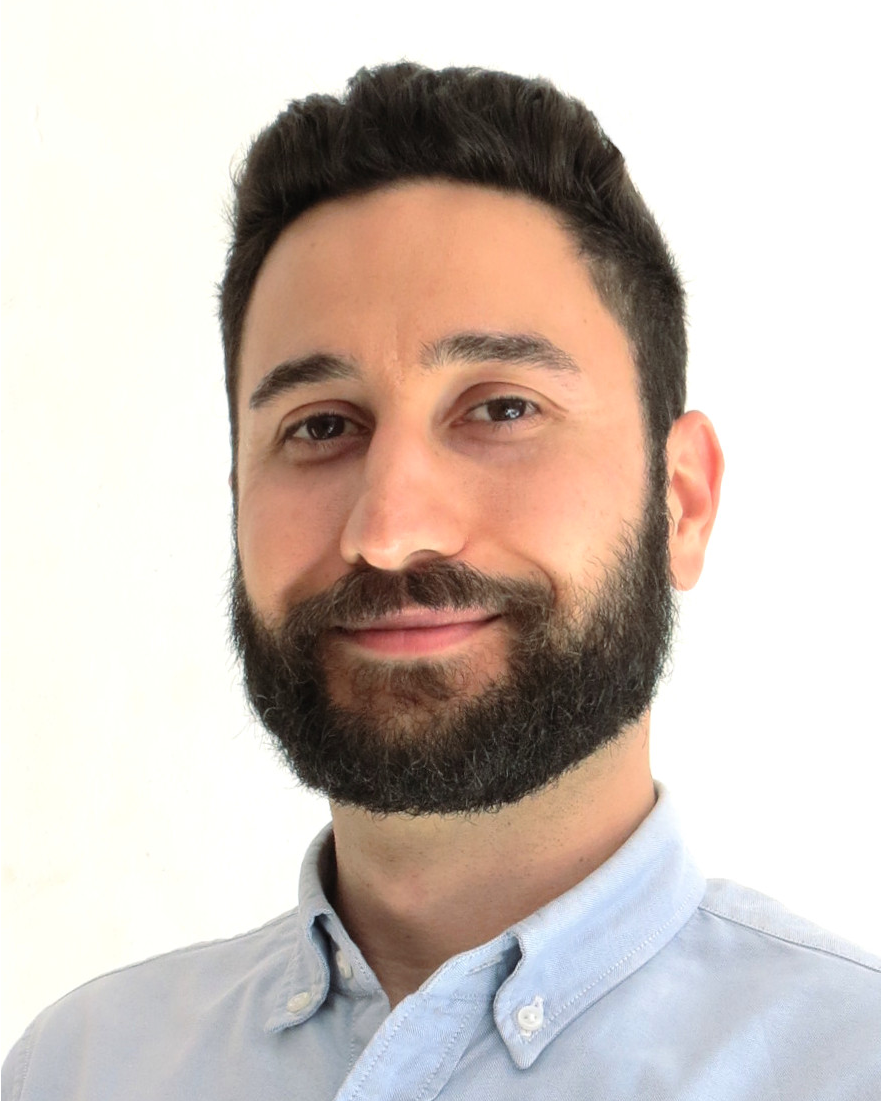}}]{Serkan Sulun}
received the Ph.D. degree in electrical and computer engineering from the University of Porto, Porto, Portugal, in 2025. He is currently a researcher with the Institute for Systems and Computer Engineering, Technology and Science (INESC TEC), Porto, Portugal. His research interests include multimedia signal processing and machine learning, with a focus on MIDI, audio, image, and video processing using deep neural networks.
\end{IEEEbiography}

\begin{IEEEbiography}[{\includegraphics[width=1in,height=1.25in,clip,keepaspectratio]{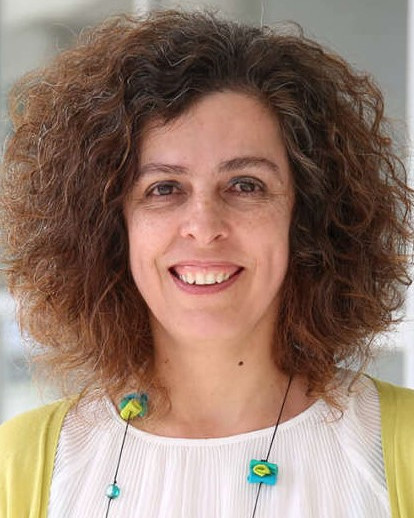}}]{Paula Viana}
(Senior Member, IEEE) received the Ph.D. degree in Electrical and Computer Engineering from the University of Porto, in 2008. She is currently a Coordinator Professor with the School of Engineering, Polytechnic of Porto, and the Head of the Multimedia Communication Technologies at INESC TEC.  She has been coordinating the participation of INESC TEC in several national and European projects. She is the author of several publications, an active reviewer for journals, conferences, and European and Portuguese research projects. Her research interests include multimedia content analysis, multimedia metadata, computer vision, multimodal machine learning and data visualisation. 
\end{IEEEbiography}

\begin{IEEEbiography}[{\includegraphics[width=1in,height=1.25in,clip,keepaspectratio]{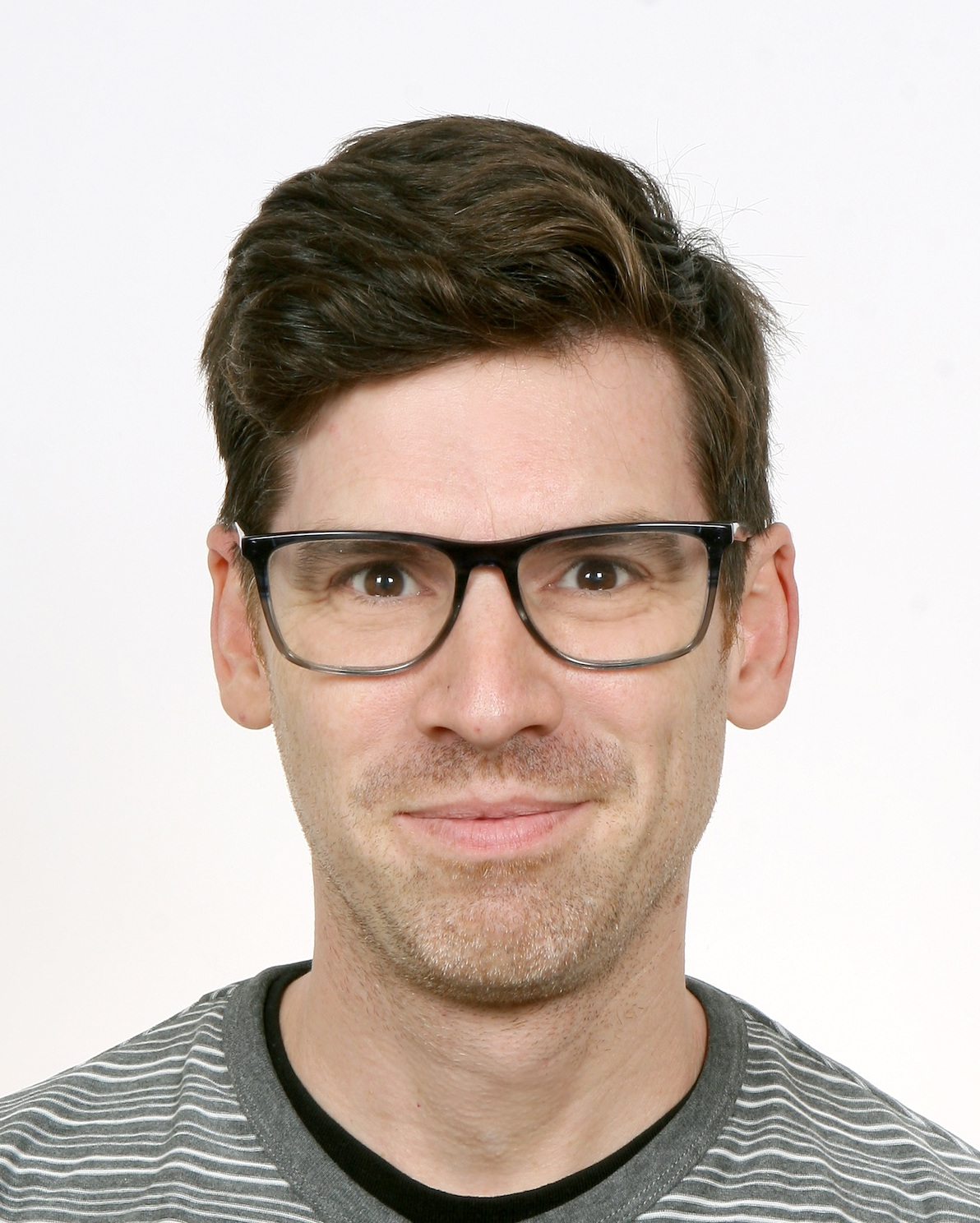}}]{Matthew E. P. Davies}
received the B.Eng. degree in computer systems with electronics from King’s College London, U.K., in 2001 and the Ph.D. degree in electronic engineering from Queen Mary University of London, U.K., in 2007. From 2007 until 2011, he was a post-doctoral researcher in the Centre for Digital Music, QMUL. In 2013, he worked in the Media Interaction Group, National Institute of Advanced Industrial Science and Technology (AIST). He coordinated the Sound and Music Computing Group at INESC TEC from 2014 to 2019, and was a researcher at the Centre for Informatics and Systems of the University of Coimbra (CISUC) from 2020 to 2021. Since 2022, he has been working in industry.
\end{IEEEbiography}

\vfill

\end{document}